\documentclass[a4paper,11pt]{article}
\usepackage{pos}

\newcommand{\beq}{\begin{equation}}
\newcommand{\eeq}{\end{equation}}
\newcommand{\bea}{\begin{eqnarray}}
\newcommand{\eea}{\end{eqnarray}}

\def\dd{{\rm d}}

\newcommand{\bem}{\begin{multline}}
\newcommand{\eem}{\end{multline}}
\newcommand{\beg}{\begin{gather}}
\newcommand{\eeg}{\end{gather}}

\newcommand{\ben}{\begin{eqnarray*}}
\newcommand{\een}{\end{eqnarray*}}

\newcommand{\bal}{\begin{align}}
\newcommand{\eal}{\begin{align}}

\newcommand{\secn}[1]{Section~1}
\newcommand{\appn}[1]{Appendix~1}

\long\def\comment#1{ }

\def\and{\quad\text{and}\quad}

\newcommand{\rme}{{\rm e}}

\def\0{{\boldsymbol 0}}

\def\l{{\boldsymbol l}}

\def\max{{\rm max}}

\title{Dynamical grooming at work: from p+p to Pb+Pb}
\author*[a]{Alba Soto-Ontoso}

\affiliation[a]{Physics Department, Brookhaven National Laboratory,\\
 Bldg. 510A, Upton, NY 11973, USA}

\emailAdd{ontoso@bnl.gov}

\abstract{We present a quantitative comparison between two grooming techniques: SoftDrop (SD) and Dynamical Grooming (DyG), in terms of the resilience of their associated observables to the underlying event, hadronization and pileup in proton-proton collisions at $\sqrt s\!=\! 13$~TeV. For that purpose, we study the groomed momentum fraction, $z_g$, in QCD jets along with boosted W tagging. In addition, we obtain the groomed momentum angle, $R_g$, within several jet quenching Monte-Carlo models in order to asses the sensitivity of SD and DyG observables to quark-gluon plasma (QGP) effects in heavy-ion collisions.}

\FullConference{%
  HardProbes2020\\
  1-6 June 2020\\
  Austin, Texas}

\begin{document}
\maketitle
The inner structure of jets provides a multi-scale probe of Quantum Chromodynamics (QCD) both in vacuum and in the presence of a hot and dense medium, such as the QGP. First-principles calculations using perturbative QCD (pQCD) together with precise measurements of jet substructure observables are instrumental to address a broad array of long-standing questions in high-energy physics. This includes the existence of new particles beyond the Standard Model, the underlying mechanism for hadronization or the nature of jet quenching, among others. In this context, a key element to enable meaningful theory-to-data comparisons are so-called grooming techniques. Their aim is to mitigate the impact of spurious radiation produced by non-perturbative phenomena, such as hadronization, underlying event or pileup, on jet substructure observables. Typically, this type of emissions populate the soft and wide-angle region of the radiation phase space. By reducing the sensitivity to this kinematic regime, grooming algorithms result into theoretically well behaved observables, i.e. systematically computable in pQCD, that might be confronted with data. 

Since the seminal work by Butterworth et al.~\cite{Butterworth:2008iy}, the catalog of grooming methods has significantly expanded over the last decade triggered by the increase in luminosity and energy of hadronic colliders~\cite{Larkoski:2017jix,Asquith:2018igt}. The preferred choice by the experimental collaborations at the LHC is SoftDrop~\cite{Larkoski:2014wba}, although ATLAS also uses trimming~\cite{Krohn:2009th}. The SoftDrop algorithm, an extension of the modified Mass Drop Tagger~\cite{Dasgupta:2013ihk}, consists in identifying the first splitting in the hardest branch of a Cambridge/Aachen (C/A)~\cite{Dokshitzer:1997in} reclustered jet that satisfies:
\begin{equation}
\label{eq:sd}
\text{SoftDrop condition:} \quad z > z_{\rm cut} \theta^\beta.
\end{equation}
In Eq.~(\ref{eq:sd}), $(z, \theta)$ correspond to the momentum sharing fraction and the angular separation of the splitting, while ($z_{\rm cut},\beta$) are free parameters with which the degree of grooming can be adjusted. All emissions with angles larger than the SD splitting are removed from the jet. 

From a theoretical point of view, the freedom to choose different combinations of ($z_{\rm cut},\beta$) in Eq.~(\ref{eq:sd}) paves the way to engineered observables particularly sensitive to a certain kinematic region of interest. However, this flexibility is disadvantageous when applying the SoftDrop technique to experimental data. In fact, there is no optimal choice of values for ($z_{\rm cut},\beta$) that, by construction, guarantees a minimization of the impact of non-perturbative radiation on different jet substructure observables. Actually, the parameters in the SD grooming condition are calibrated with Monte-Carlo simulations on an observable-by-observable basis. This fine tuning exercise is nicely exemplified in a new ATLAS conference note~\cite{ATLAS:2020zhd}, where an optimal jet reconstruction scheme is investigated by combining grooming and pileup mitigation methods. The situation is even more cumbersome in heavy-ion collisions, where no well-established jet quenching Monte-Carlo exists. Traditionally, the values of ($z_{\rm cut},\beta$) tailored for jet studies in p+p~\cite{Adam:2020kug,Aad:2019vyi} have been adopted by heavy-ion analyses~\cite{Acharya:2019djg,Sirunyan:2017bsd,Kauder:2017mhg}. It was only recently that dedicated studies addressed the rather different challenges that grooming algorithms have to face in a heavy-ion environment~\cite{Mulligan:2020tim}. 

At this point, it's natural to wonder whether the amount of unconstrained parameters in grooming algorithms might be reduced. To that end, a novel family of so-called 'dynamical groomers' has been proposed in~\cite{Mehtar-Tani:2019rrk,Mehtar-Tani:2020oux}. The core idea is to use the 'hardest' emission inside the jet to define a fluctuating grooming condition that is auto-generated on a jet-by-jet basis. More concretely, one looks for the splitting along the primary Lund plane~\cite{Andersson:1988gp,Dreyer:2018nbf} of the C/A reclustered jet that satisfies:

\begin{equation}
\label{eq:dyg}
\text{Dynamical Grooming condition:} \quad \kappa^{(a)} =\frac{1}{p_T}\,\,\underset{i\in\, \text{LP}}{\max}\left[z_i(1-z_i) \, p_{T,i}\,\left(\frac{\theta_i}{R}\right)^a  \right],
\end{equation}
where the $(1-z_i)$ factor is introduced such that $\kappa^{(a)}$, dubbed 'hardness', is a relative quantity with respect to the parent emitter. In addition, $a$ in Eq.~(\ref{eq:dyg}) is a positive (to ensure IRC safe observables) and continuous free parameter that defines the meaning of 'hardest' emission, e.g. $a\!=\!1$ corresponds to select the splitting with the largest relative $k_T$. After identifying the hardest splitting, branches located prior in the C/A sequence are groomed away. 

Conceptually, the major difference between DyG and existing grooming methods concerns the nature of the scale at which the splittings phase-space is suppressed. Notice that in SD, and other techniques such as trimming, this scale is a sharp cut-off given as an input to the algorithm (see Eq.~(\ref{eq:sd})). This is in stark contrast to the dynamically generated cut-off in DyG that scales as $\sim\!\sqrt{\alpha_s/a}$. Then, the main advantage of all these new, dynamical variables given by Eq.~(\ref{eq:dyg}) is that the strong coupling itself governs the strength of the grooming.

The building block to analytically compute DyG observables is the two-dimensional probability distribution for a splitting to be the hardest in an angular ordered shower, i.e.
\begin{equation}
\label{eq:prob-dist}
{\cal P}(z,\theta) = \frac{\alpha_s(k^2_T)}{\pi} \, zP(z)\, \Delta(\kappa \big| a)  \,,
\end{equation}
where $P(z)$ is the splitting function and $\Delta(\kappa \big| a)$ is a Sudakov form factor given by
\begin{equation}
\label{eq:sudakov-0}
\Delta \big(\kappa \big|a)= \exp \left[-  \int_0^{R} \frac{\dd \theta}{\theta} \int_0^1 \dd z\,\frac{\alpha_s(k^2_t)}{\pi} P(z)\times \Theta \big( z(1-z) (\theta/R)^a > \kappa \big)\right].
\end{equation}
The Heaviside function in Eq.~(\ref{eq:sudakov-0}) enforces all emissions to be softer than the tagged one, i.e. to have a smaller value of $\kappa^{(a)}$. In this way, the properties of the hardest emission can be accessed within pQCD from Eq.~(\ref{eq:prob-dist}), up to the required logarithmic precision~\cite{Mehtar-Tani:2019rrk}.
 \begin{figure}
\centering
\includegraphics[width=0.33\textwidth]{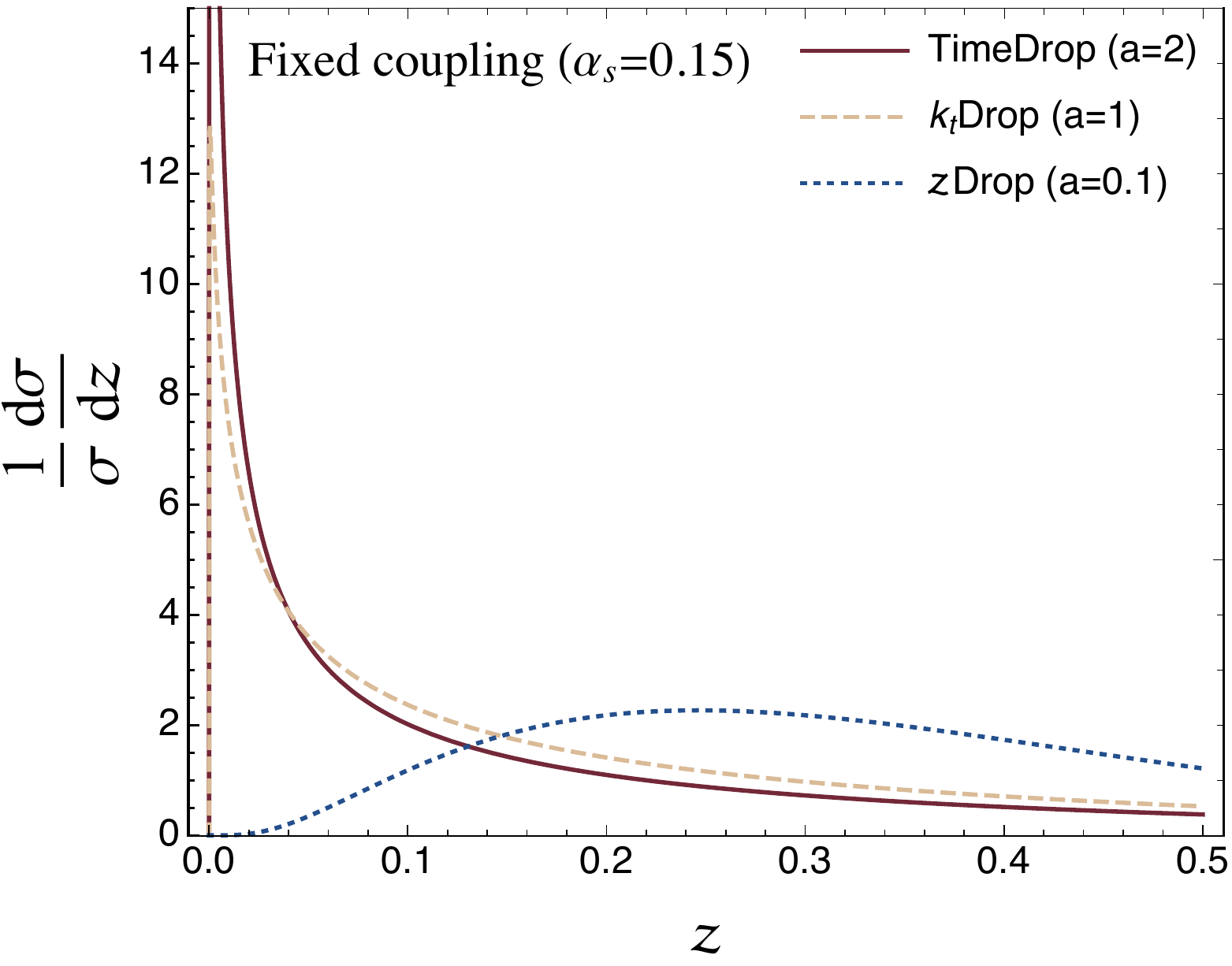}\quad \includegraphics[width=0.3\textwidth]{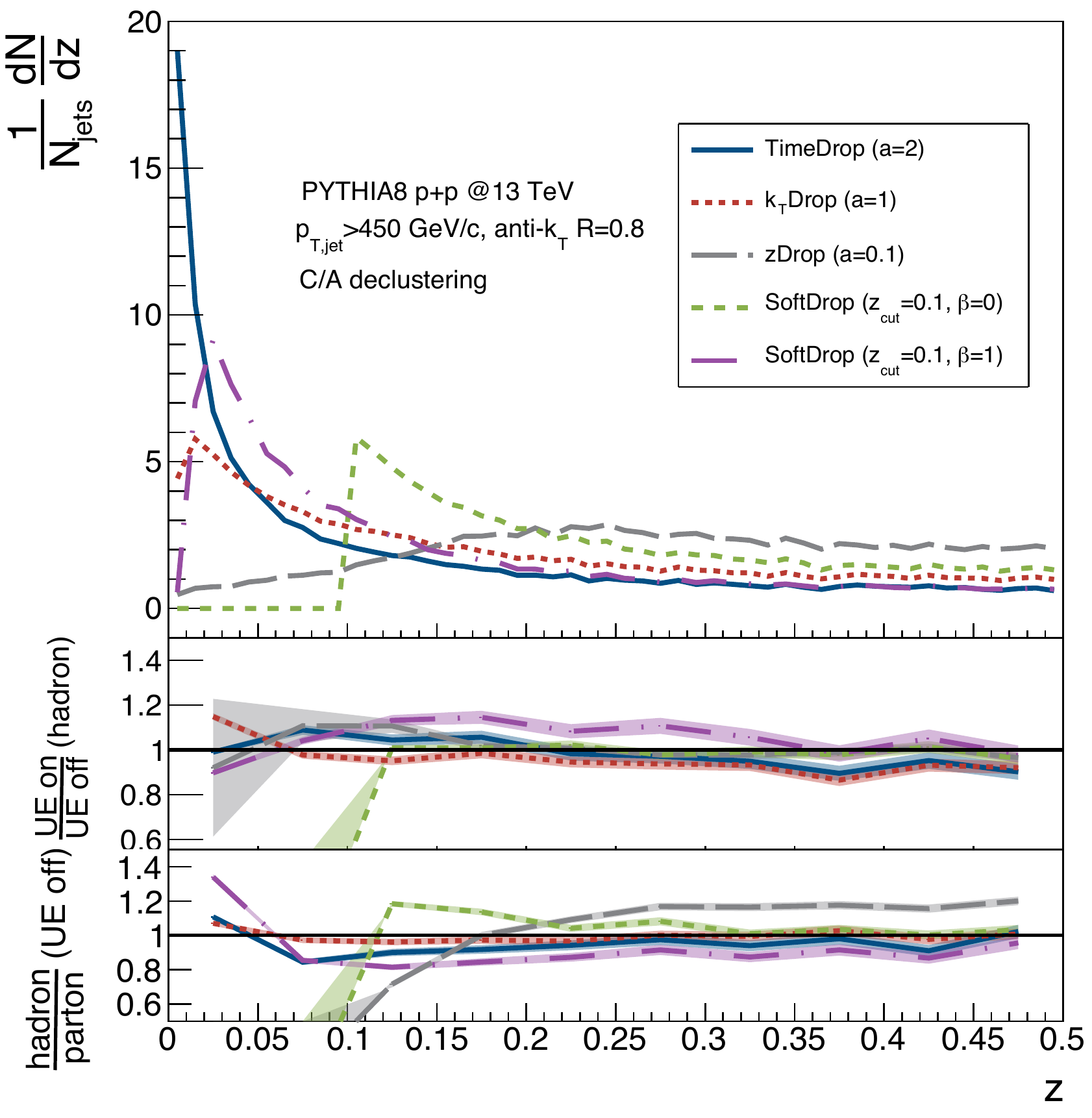}\quad \includegraphics[width=0.3\textwidth]{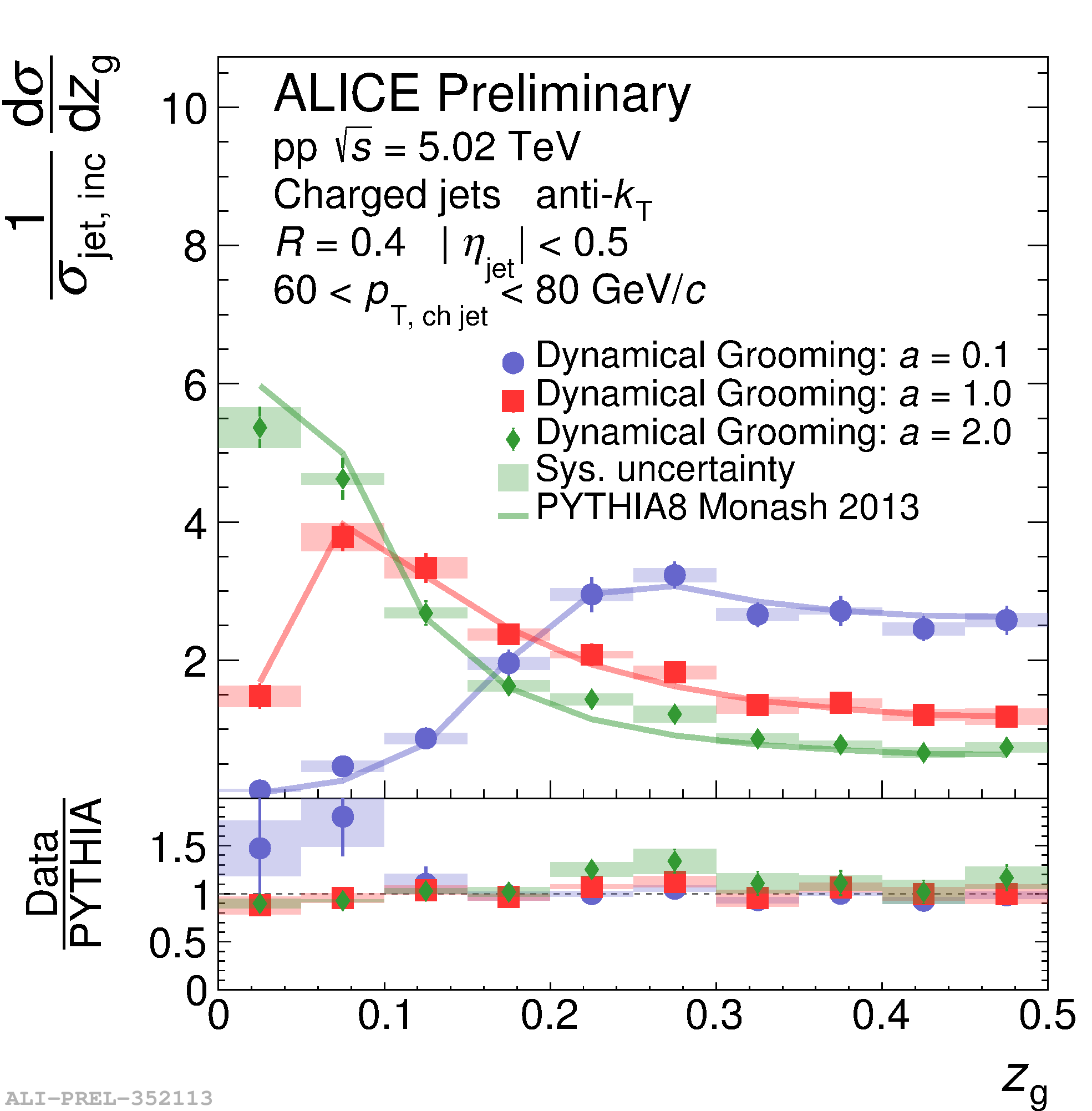}
\caption{Dynamically groomed $z_g$ distribution. Left: analytic result at modified leading-log accuracy with fixed coupling. Middle: Monte-Carlo study including underlying event and hadronization effects. Right: ALICE preliminary data~\cite{Mulligan:2020}. }
\label{fig:zg}
\end{figure}

\begin{figure}
\centering
\includegraphics[width=0.3\textwidth]{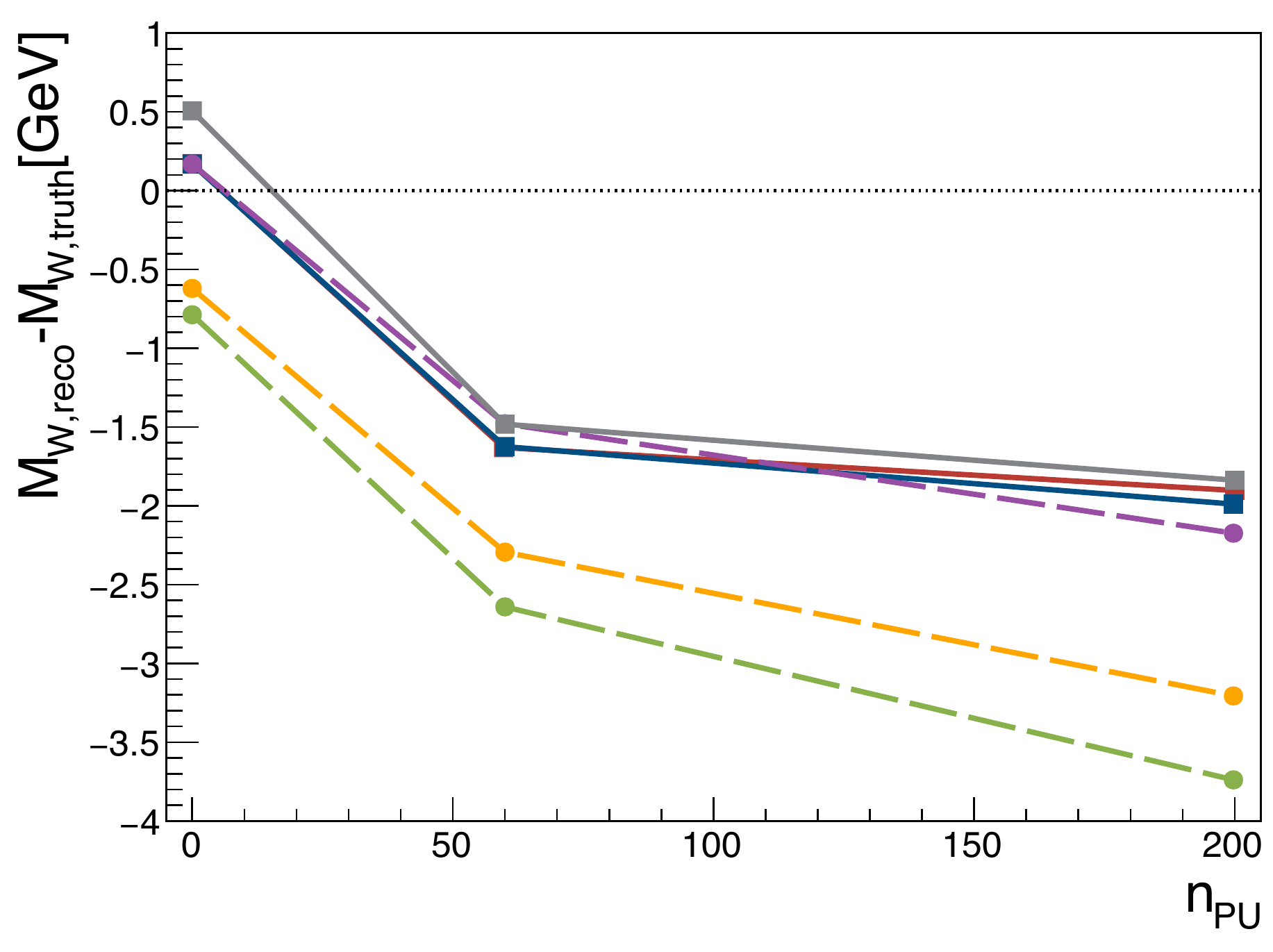}\quad \includegraphics[width=0.31\textwidth]{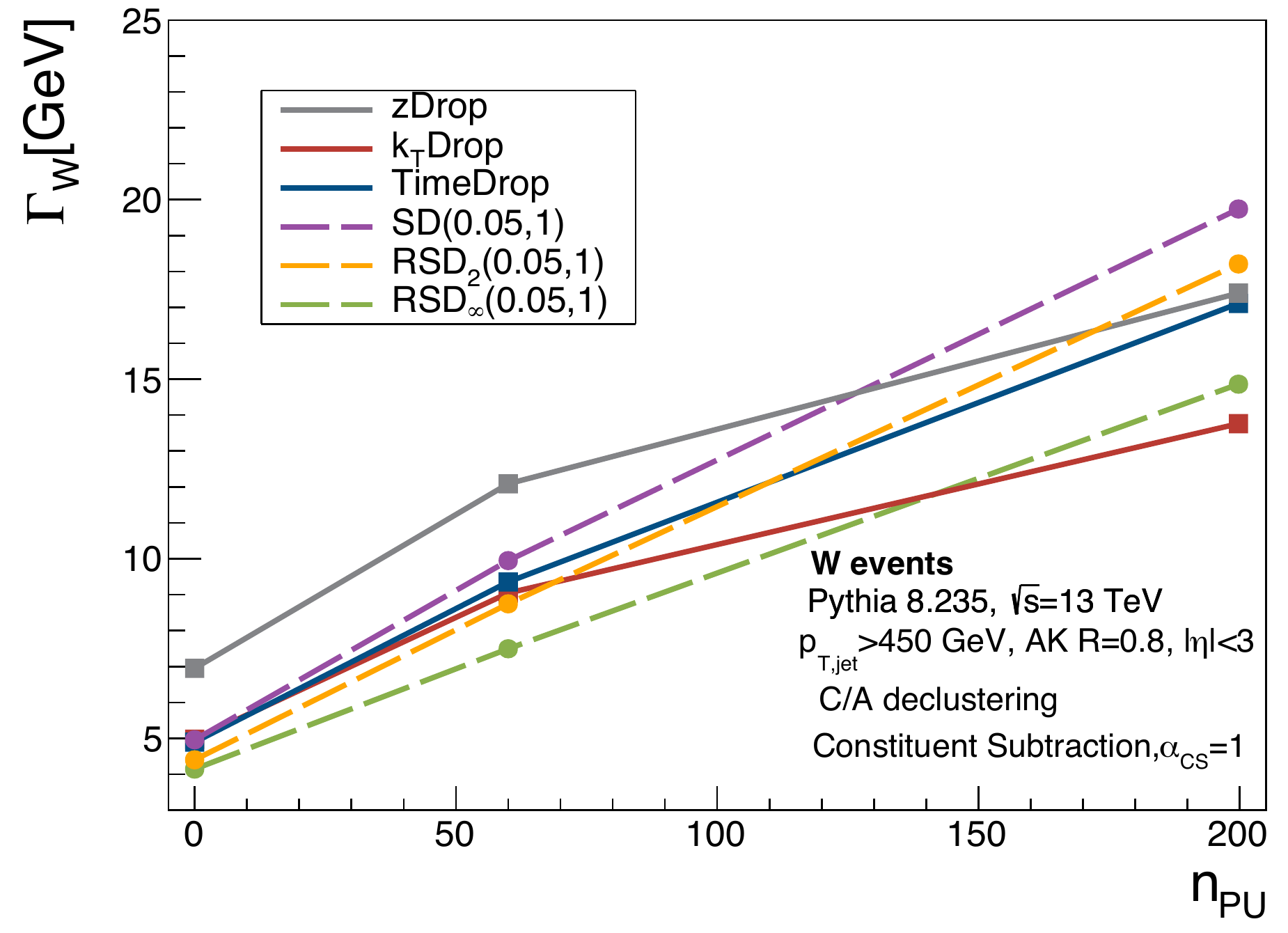}\quad \includegraphics[width=0.32\textwidth]{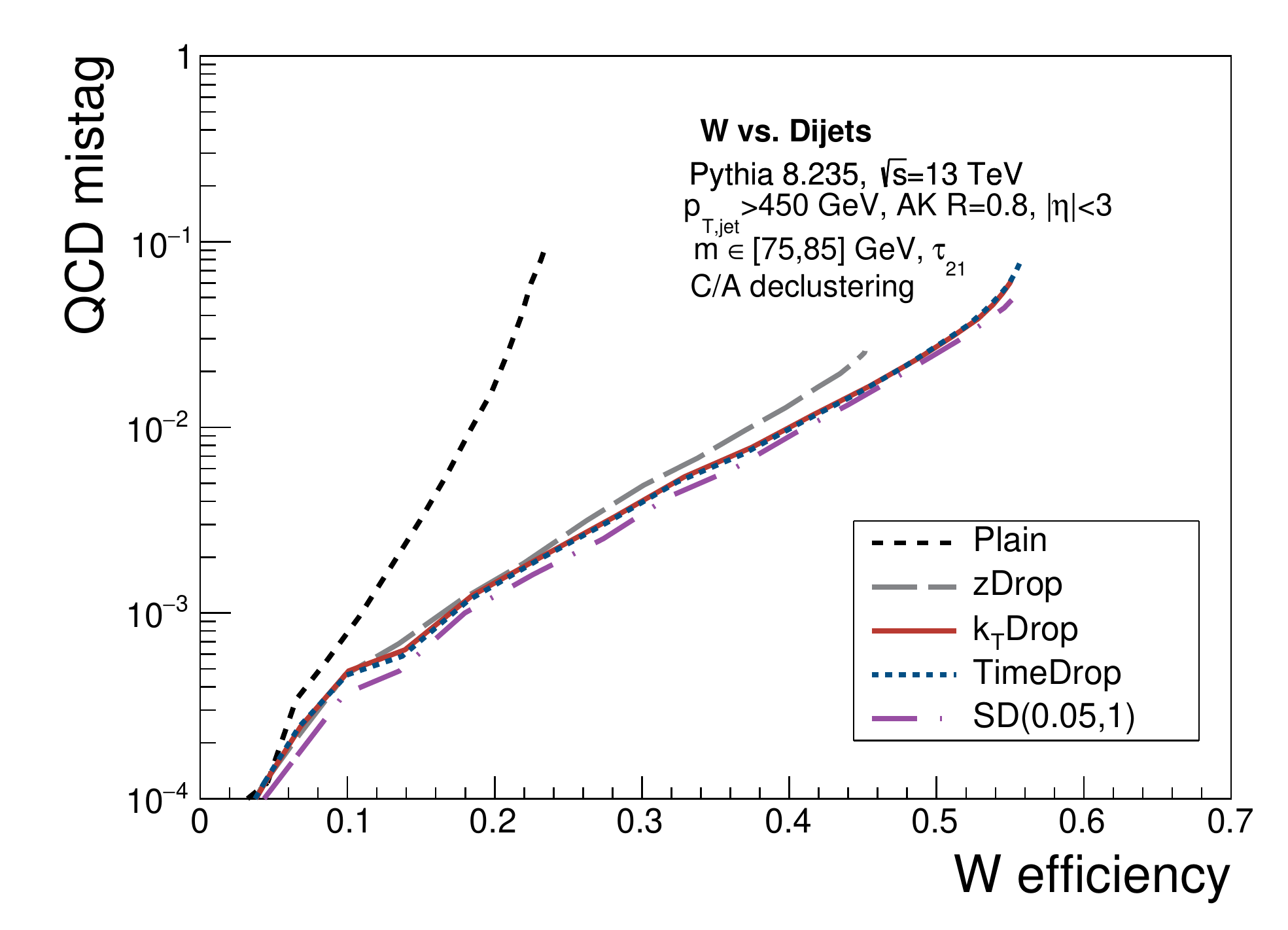}
\caption{W mass scale (left) and resolution (middle) as a function of the number of pileup interactions with different grooming methods. Right: ROC curve for W vs. dijets discrimination.}
\label{fig:w-tagging}
\end{figure}
As a working example, the momentum sharing fraction  of the tagged splitting, known as $z_g$, can be obtained by integrating the angular dependence of ${\cal P}(z,\theta)$. The resulting distribution is shown on the left panel of Fig.~\ref{fig:zg} for different choices of $a$ in Eq.~(\ref{eq:dyg}). A very distinctive property of these distributions is their visible drop-off at $z_{\rm cut} \approx \rme^{-\sqrt{a/\alpha_s}}$. Note that this cut-off is dynamically generated and not a pre-fixed value as in SD when $\beta\!=\!0$. Furthermore, when $z\!>\!z_{\rm cut}$ the DyG distributions are proportional to the splitting function, modulated by a factor $\sqrt{\alpha_s/a}$, thus providing a way to experimentally access $P(z)$ in a broad range of $z$ values. These features are confirmed by PYTHIA8~\cite{Sjostrand:2007gs} simulations of di-jet events at $\sqrt s\!=\!13$~TeV, as shown on the middle panel of Fig.~\ref{fig:zg} where a couple of SoftDrop settings are also presented\footnote{Details on the jet reconstruction procedure are given as labels on the plots along this paper.}. On the lower panels of this plot, the sensitivity of this observable to the underlying event (top) and hadronization (bottom) is tested. Regarding the underlying event, we observe that DyG behaves similarly, overall, to SoftDrop, specially when $\beta\!=\!0$. The differences among the compared methods are highlighted when focusing on the hadronization impact. In particular, $k_T$Drop ($a\!=\!1$ in Eq.~(\ref{eq:dyg})) shows an enhanced resilience to this non-perturbative phenomenon over the whole range of $z$. This fact can be understood as a result of suppressing radiation with $k_T\!<\!\lambda_{\rm QCD}$. Finally, ALICE has measured for the first time the DyG $z_g$-distribution in p+p collisions at $\sqrt s\!=\!5.02$~TeV. As can be seen in the right panel of Fig.~\ref{fig:zg}, the data is well-described by PYTHIA8 and in qualitative agreement with our analytic estimates. 
 
To further illustrate the differences and similarities in the performance of SD and DyG, we resort to a traditional test case for grooming methods: boosted W tagging. To that end, we generate $p\!+\!p\!\rightarrow\!WW$ events enforcing $W\!\rightarrow\!q\bar{q}$ and obtain the peak, $M_{W,{\rm reco}}$, and the width, $\Gamma_{W,}$,  of the reconstructed $W$ mass distribution for different pileup conditions characterized by the number of minimum bias events, $n_{\rm PU}$, in which the signal is embedded. It's worth noting that when pileup is included, any grooming method has to be supplemented with a dedicated pileup mitigation algorithm in order to recover a realistic mass distribution. Hence, we apply Constituent Subtraction on a jet-by-jet level and then groom. The values of $M_{W,{\rm reco}}$ and $\Gamma_{W,}$ are laid out on the left and middle panels, respectively, of Fig.~\ref{fig:w-tagging} for no pileup, current LHC ($n_{\rm PU}\!=\!60$) and HL-LHC ($n_{\rm PU}\!=\!200$). We compare dynamical groomers with SD and its recursive generalization (RSD) with fine-tuned parameters for this concrete observable~\cite{Dreyer:2018tjj}. In the case of $M_{W,{\rm reco}}$, we find a trend to over-subtract with increasing pileup, particularly enhanced in the case of Recursive SoftDrop, that might be suffocated by exploring other pileup mitigation settings. Regarding the width of the W mass distribution, we observe a remarkable robustness against the number of pileup interactions for Dynamical Grooming with $a\!=\!1$. It's worth emphasizing that this narrowing in the distribution doesn't come at the cost of a significant shift in the peak position as it is the case for RSD (see left panel of Fig.~\ref{fig:w-tagging}). All in all, $k_T$Drop gives the best performance on boosted W tagging.

In order to tag the hadronically decaying boosted W is not enough to reconstruct the mass distribution accurately, but also to distinguish it from a QCD jet that falls in the same mass bin. Therefore, other jet properties such as the expected number of hard prongs (1 for QCD jets, 2 for W jets) might be exploited in order to perform such a task. Following this idea, we select the N-subjettiness ratio $\tau_{21}$~\cite{Thaler:2010tr} as the second discriminatory variable in our analysis. The ability of different groomers to distinguish a W jet from a QCD one is quantified via ROC curves and displayed on the right panel of Fig.~\ref{fig:w-tagging}. First, we observe how grooming leads to a factor $\!\sim\!3$ improvement on the W tagging efficiency in the worst QCD mistag scenario. Then, except for $z$Drop, a quantitative agreement between the SD result and out-of-the-box dynamical groomers is demonstrated thus confirming this new method as a robust tool for experimental jet substructure analyses.  
 \begin{figure}
\centering
\includegraphics[width=0.8\textwidth]{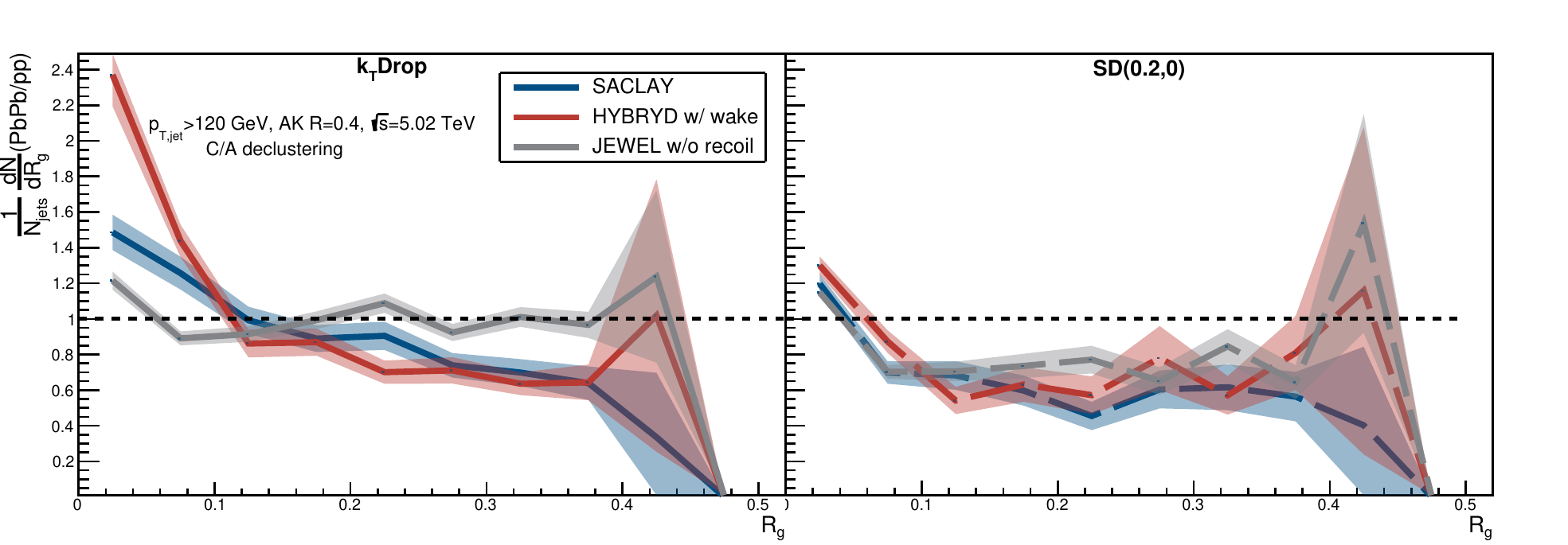}
\caption{The PbPb/pp ratio of the $R_g$ distribution obtained with Dynamical Grooming (left) and the ALICE-like SoftDrop setting (right) for different jet quenching Monte-Carlo models: Saclay~\cite{Caucal:2019uvr}, Hybrid~\cite{Casalderrey-Solana:2014bpa} and JEWEL~\cite{Zapp:2008gi}. }
\label{fig:rg}
\end{figure}

Up to now, we have quantified the performance of DyG in p+p like observables. A natural extension would be jet substructure studies in heavy-ion collisions. One of the main questions that we would like to address is whether DyG observables have a larger discriminating power than conventional SoftDrop ones. A first step towards this goal is displayed in Fig.~\ref{fig:rg}, where we perform a preliminary, Monte Carlo study of the ratio of the $R_g$-distribution in Pb+Pb collisions with respect to p+p using DyG with $a\!=\!1$ and SD with ($z_{\rm cut},\beta$) as chosen in the recent ALICE measurement. The chosen Monte-Carlo models have certain similarities in the parton shower treatment, but clearly differ on the medium description, e.g. the Hybrid result includes medium response while Saclay's treats the QGP as a brick of fixed length. Interestingly, these differences are amplified in the $R_g$-distribution when computed with DyG with respect to the SD setting. This promising result opens up the possibility to deepen our understanding of parton energy loss by means of these new class of jet substructure observables. \section*{Acknowledgements}
The author would like to thank Yacine Mehtar-Tani and Konrad Tywoniuk for their collaboration in developing dynamical grooming. This work was supported by the U.S. Department of Energy, Office of Science, Office of Nuclear Physics, under contract No. DE- SC0012704,
and by Laboratory Directed Research and Development (LDRD) funds from Brookhaven Science Associates.

\bibliographystyle{apsrev4-1}
\bibliography{dyg}

\end{document}